\documentclass{egpubl}

\JournalSubmission    
\electronicVersion 

\ifpdf \usepackage[pdftex]{graphicx} \pdfcompresslevel=9
\else \usepackage[dvips]{graphicx} \fi

\PrintedOrElectronic

\usepackage{t1enc,dfadobe}
\usepackage{lineno}
\usepackage{egweblnk}
\usepackage{cite}
\usepackage{amsmath}
\usepackage{subcaption}
\usepackage{float}
\usepackage{amssymb}

\usepackage{xspace}
\title[Progressive Transient Photon Beams]%
      {Progressive Transient Photon Beams}

\author[Marco et al.]
{
	\parbox{\textwidth}{\centering Julio Marco$^{1}$ \qquad Ib\'{o}n Guill\'{e}n$^{1}$ \qquad Wojciech Jarosz$^{2}$ \qquad Diego Gutierrez$^{1}$ \qquad Adrian Jarabo$^{1}$ }
        \\
{\parbox{\textwidth}{\centering $^{1}$ Universidad de Zaragoza, I3A \qquad $^{2}$ Dartmouth College}
}}

\begin{document}
\graphicspath{{./figures/}}

\newcommand{\fref}[1]{Figure~\ref{#1}}
\newcommand{\tref}[1]{Table~\ref{#1}}
\newcommand{\eref}[1]{Equation~\ref{#1}}
\newcommand{\cref}[1]{Chapter~\ref{#1}}
\newcommand{\sref}[1]{Section~\ref{#1}}
\newcommand{\aref}[1]{Annex~\ref{#1}}

\definecolor{red}{rgb}{0.8,0,0}
\definecolor{purered}{rgb}{1,0,0}
\definecolor{darkred}{rgb}{0.6,0,0}
\definecolor{green}{rgb}{0.0,0.5,0}
\definecolor{blue}{rgb}{0,0,0.75}
\definecolor{darkblue}{rgb}{0,0,0.55}
\definecolor{orange}{rgb}{0.9,0.3,0.1}
\definecolor{purple}{rgb}{0.6,0.0,0.6}
\definecolor{cyan}{rgb}{0.0,0.7,0.7}
\definecolor{darkgray}{rgb}{0.4,0.4,0.4}
\definecolor{bronze}{rgb}{0.8, 0.5, 0.2}
\definecolor{dorange}{rgb}{0.75, 0.4, 0.0}

\newcommand{\julioc}[1]{\textcolor{blue}{\emph{(\textbf{Julio:} #1)}}}
\newcommand{\juliotxt}[1]{\textcolor{blue}{\emph{\textbf{Julio:} #1}}}
\newcommand{\julio}[1]{{\leavevmode\color{blue}{#1}}}
\newcommand{\ibonc}[1]{\textcolor{orange}{\emph{(\textbf{Ibon:} #1)}}}
\newcommand{\adrianc}[1]{\textcolor{purple}{\emph{(\textbf{Adrian:} #1)}}}
\newcommand{\forum}[1]{{\leavevmode\color{black}{#1}}}
\newcommand{\corrected}[1]{\textcolor{dorange}{#1}}

\newcommand*{\pmapping}[0]{\emph{photon mapping}}
\newcommand*{\Pmapping}[0]{\emph{Photon mapping}}
\newcommand*{\PMapping}[0]{\emph{Photon Mapping}}
\newcommand*{\pbeams}[0]{\emph{photon beams}}
\newcommand*{\Pbeams}[0]{\emph{Photon beams}}
\newcommand*{\PBeams}[0]{\emph{Photon Beams}}
\newcommand*{\Ptracing}[0]{\emph{Path tracing}}
\newcommand*{\VPTracing}[0]{\emph{Volumetric Path Tracing}}
\newcommand*{\PTracing}[0]{\emph{Path Tracing}}
\newcommand*{\ptracing}[0]{\emph{path tracing}}
\newcommand*{\Bptracing}[0]{\emph{Bidirectional path tracing}}
\newcommand*{\bptracing}[0]{\emph{bidirectional path tracing}}
\newcommand*{\Metropolis}[0]{\emph{Metropolis light transport}}
\newcommand*{\brdf}[0]{\emph{BRDF}}
\newcommand*{\ns}[0]{\mathrm{ns}}

\newcommand{\Real}{\mathbb{R}}
\newcommand{\Tr}{{T_r}}
\newcommand{\abs}{\mu_a} 
\newcommand{\sca}{\mu_s} 
\newcommand{\ext}{\mu_t} 
\newcommand{\alb}{{_\Lambda}}
\newcommand{\scalb}{\alpha}
\newcommand{\crossscat}{\kappa_s}
\newcommand{\crossabs}{\kappa_a}
\newcommand*{\pf}[0]{\rho}
\newcommand*{\coefext}[0]{\sigma_t}
\newcommand*{\coefextx}[0]{\ext(x)}
\newcommand*{\coefabs}[0]{\sigma_a}
\newcommand*{\coefabsx}[0]{\abs(x)}
\newcommand*{\coefscat}[0]{\sigma_s}
\newcommand*{\coefscatx}[0]{\sca(x)}
\newcommand*{\mcoefext}[0]{$\coefext$}
\newcommand*{\mcoefextx}[0]{$\coefextx$}
\newcommand*{\mcoefabs}[0]{$\coefabs$}
\newcommand*{\mcoefabsx}[0]{$\coefabsx$}
\newcommand*{\mcoefscat}[0]{$\coefscat$}
\newcommand*{\mcoefscatx}[0]{$\coefscatx$}
\newcommand*{\phasefunction}[0]{p(x, \vec{\omega}^\prime, \vec{\omega})}
\newcommand*{\ior}[0]{\eta}
\newcommand{\diff}{\mathrm{d}}

\newcommand{\sDeltaTime}{\tau}

\newcommand{\omegav}{\vec{\omega}}
\newcommand{\omegaout}{\omegav_o}
\newcommand{\omegain}{\omegav_i}
\newcommand{\momegain}{$\omegain$}
\newcommand{\ppwr}{\Phi}
\newcommand{\eqbreak}{\nonumber \\}
\newcommand{\x}{\ensuremath{\mathbf{x}}\xspace}
\newcommand{\y}{\ensuremath{\mathbf{y}}\xspace}
\newcommand{\z}{\ensuremath{\mathbf{z}}\xspace}
\newcommand{\rvec}{\vec{\ensuremath{\mathbf{r}}}\xspace}
\newcommand{\rlen}{\ensuremath{\mathrm{r}}\xspace}
\newcommand{\xpr}{\ensuremath{\mathbf{x}^\prime}\xspace}
\newcommand{\xm}{{\x}}
\newcommand{\xc}{{\x_r}}
\newcommand{\xb}{{\xm_b}}
\newcommand{\ym}{{\y}}
\newcommand{\zm}{{\z}}
\newcommand{\VRegion}{\ensuremath{\aleph}\xspace}
\newcommand{\xs}{\ensuremath{\x_{s}}\xspace}
\newcommand{\ys}{\ensuremath{\y_{\! s}}\xspace}
\newcommand{\xw}{\ensuremath{\hat{\x}}\xspace}
\newcommand{\yw}{\ensuremath{\hat{\y}}\xspace}
\newcommand{\mpar}{q}
\newcommand{\pkern}{\Omega}
\newcommand{\optpath}{\Pi}


\newcommand{\sPixel}{L}             
\newcommand{\sPixelSample}{\widehat{\sPixel}}             
\newcommand{\estimate}[1]{\ensuremath{\widehat{#1}}}
\newcommand{\sTime}{t}
\newcommand{\sK}{\ensuremath{K}}
\newcommand{\sR}{\ensuremath{R{_b}}}
\newcommand{\sKoneD}{\ensuremath{\sK_{\textrm{1D}}}}
\newcommand{\sT}{\ensuremath{\mathcal{T}}}
\newcommand{\sKt}{\ensuremath{\sK_\sT}}
\newcommand{\sEV}[1]{\ensuremath{\textrm{E}[#1]}}
\newcommand{\sError}{\ensuremath{\epsilon}}
\newcommand{\sVar}[1]{\ensuremath{\textrm{Var}[#1]}}
\newcommand{\Order}[1]{\ensuremath{O(#1)}}
\newcommand{\order}[1]{\ensuremath{\Theta(#1)}}
\newcommand{\Lo}{L} 
\newcommand{\Beta}{\ensuremath{\textrm{B}}}
\newcommand{\AMSE}{\ensuremath{\textrm{AMSE}}}

\newcommand{\sEyeContribution}{\ensuremath{\Psi}}
\newcommand{\sLightContribution}{\ensuremath{\ppwr}}

\newcommand{\sRandomVariableR}{\ensuremath{D}}
\newcommand{\sRandomVariableT}{\ensuremath{T}}
\newcommand{\sProbabilityRVar}[1]{\ensuremath{p_{\sR}}}
\newcommand{\sProbabilityR}{\ensuremath{\sProbabilityRVar{\x}}}
\newcommand{\sProbabilityTVar}[1]{\ensuremath{p_{\sT}}}
\newcommand{\sProbabilityT}{\ensuremath{\sProbabilityTVar{\sTime}}}

\newcommand{\sKCanonical}{\ensuremath{k}}
\newcommand{\sKoneDCanonical}{\ensuremath{\sKCanonical_{1D}}}
\newcommand{\sKtCanonical}{\ensuremath{\sKCanonical_\sT}}
\newcommand{\sKVar}{\ensuremath{\xi}}
\newcommand{\sKVarChange}{\ensuremath{\psi}}

\newcommand{\sConstantK}{\ensuremath{{\mathcal{C}}}}
\newcommand{\sConstantKoneD}{\ensuremath{{\sConstantK_{1D}}}}
\newcommand{\sConstantKt}{\ensuremath{{\sConstantK_\sT}}}
\newcommand{\sConstantSecondKoneD}{\ensuremath{{\sConstantK^{ii}_{1D}}}}
\newcommand{\sConstantSecondKt}{\ensuremath{{\sConstantK^{ii}_\sT}}}

\newcommand{\scalePar}{\ensuremath{\beta}}
\newcommand{\scaleParR}{\ensuremath{\scalePar_\sR}}
\newcommand{\scaleParT}{\ensuremath{\scalePar_\sT}}

\definecolor{red}{rgb}{0.8,0,0}
\definecolor{purered}{rgb}{1,0,0}
\definecolor{darkred}{rgb}{0.6,0,0}
\definecolor{green}{rgb}{0.0,0.5,0}
\definecolor{blue}{rgb}{0,0,0.75}
\definecolor{darkblue}{rgb}{0,0,0.55}
\definecolor{black}{rgb}{0,0,0}
\definecolor{orange}{rgb}{0.9,0.3,0.1}
\definecolor{purple}{rgb}{0.6,0.0,0.6}
\definecolor{cyan}{rgb}{0.0,0.7,0.7}
\definecolor{darkgray}{rgb}{0.4,0.4,0.4}
\definecolor{bronze}{rgb}{0.8, 0.5, 0.2}
\newcommand{\changes}[1]{#1}
\teaser{
	\centering
	\includegraphics[width=0.95\textwidth]{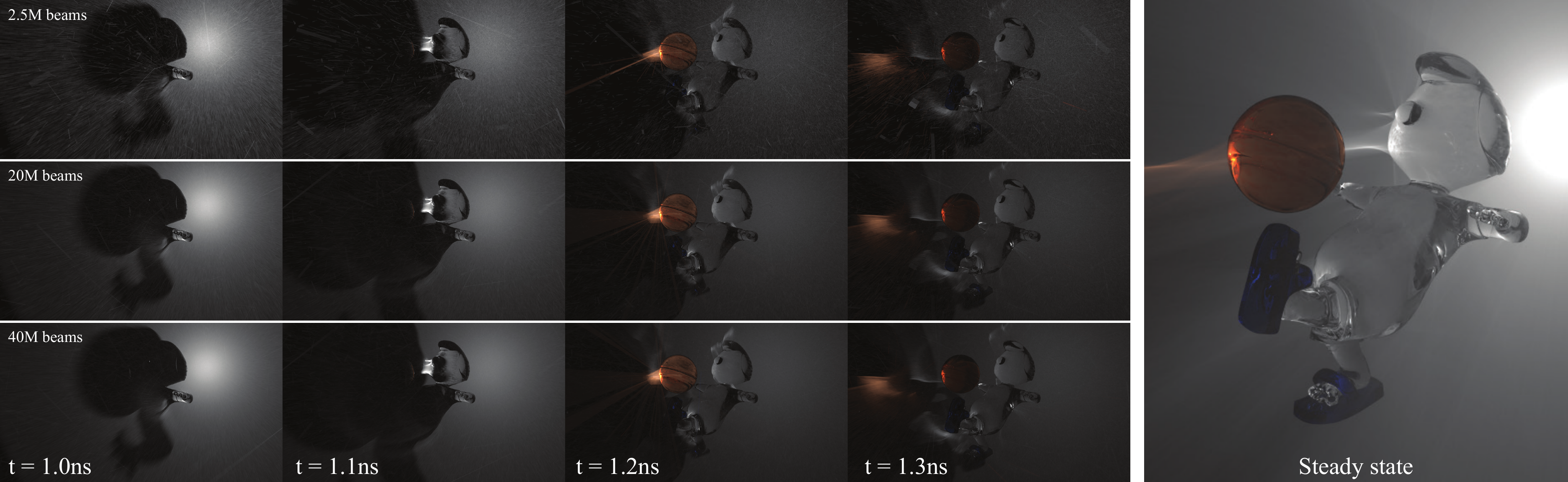}
	\caption{
		\forum{We present a robust method for transient rendering in participative media. We reformulate photon beams to support transient light propagation, and derive a progressive approach that performs spatio-temporal density estimations. Here we show different frames of transient light transport (left sequences) in the \textsc{Soccer} scene (steady-state render on the right), at different levels of convergence. The scene features complex caustic light transport in the medium due to multiple reflections and refractions in smooth dielectrics inside the medium. 
The progressive nature of our algorithm is consistent under finite memory, and works by accumulating several independent iterations of transient photon beams that progressively reduce both bias and variance. Please refer to the supplemental video for the full sequence.}}
	\label{fig:soccer}
}

\maketitle

\begin{abstract}
In this work we introduce a novel algorithm for transient rendering in participating media. Our method is consistent, robust, and is able to generate animations of time-resolved light transport featuring complex caustic light paths in media. 
We base our method on the observation that the spatial continuity provides an increased coverage of the temporal domain, and generalize photon beams to transient-state. We extend the beam steady-state radiance estimates to include the temporal domain. Then, we develop a progressive version of spatio-temporal density estimations, that converges to the correct solution with finite memory requirements by iteratively averaging several realizations of independent renders with a progressively reduced kernel bandwidth. We derive the optimal convergence rates accounting for space and time kernels, and demonstrate our method against previous consistent transient rendering methods for participating media.

\ccsdesc[300]{Computer Graphics~Three-dimensional graphics and realism}
\ccsdesc[300]{Computer Graphics~Raytracing}
\ccsdesc[300]{Computer Graphics~Transient rendering}

\printccsdesc   
\end{abstract}  

\section{Introduction}

The emergence of transient imaging has led to a vast number of applications in graphics and vision~\cite{Jarabo2017transient}, where the ability of sensing the world at extreme high temporal resolution allows new applications such as imaging light in motion~\cite{Velten2013}, appearance capture~\cite{Naik2011}, geometry reconstruction~\cite{Busck2004,MarcoSIGA2017DeepToF}, or vision through media~\cite{Busk2005underwater,Wu2018adaptive} and around the corner~\cite{Velten2012nc,Arellano2017NLOS}. 
Sensing through media is one of the key applications: The ability of demultiplexing light interactions in the temporal domain is a very promising approach for important practical domains such as non-invasive medical imaging, underwater vision, or autonomous driving through fog. Accurately simulating light transport could enormously help in this applications, potentially serving as benchmark, forward model in optimization, or as a training set for machine learning. 

Transient rendering in media is, however, still challenging: The increased dimensionality (time) essentially increase variance in Monte Carlo algorithms, which might lead to unpractical rendering times. This variance is specially harmful in media, where the signal tends to be smooth due to the low-pass filter behavior of scattering, in both the spatial and temporal domains. 
One of the major drawbacks of transient rendering is that it requires much higher sampling rates to fill up the extended temporal domain, specially when using $0D$ point samples, which are sparsely distributed along time. We make the observation that leveraging the continuity of full photon trajectories allows us to densely populate both space and time. 
The natural conclusion of that observation is that using a technique based on photon beams~\cite{Jarosz2011} should significantly reduce the rendering time when computing a noise-free time-resolved render. Moreover, given the density estimation nature of photon beams, it naturally combines with the reconstruction technique on the temporal domain proposed by Jarabo et al.~\cite{Jarabo2014}.


\forum{In this work, we present a new method for rendering participating media in transient state, that leverages the good properties of density estimation for reconstructing smooth signals. Our method extends \emph{progressive photon beams} (PPB)~\cite{jarosz2011progressive} to the time domain, and combines it with temporal density estimation for improved reconstruction in both the spatial and temporal domains. Our technique is biased, but consistent with finite memory, by taking advantage on the progressive nature of density estimation. 
Then we analyze the asymptotic convergence of our proposed space-time density estimation, computing the optimal kernel reduction ratios for both domains. Finally, we demonstrate our method on a variety of scenes with complex volumetric light transport, featuring high-frequency occlusions, caustics, or glossy reflections, and show its improved performance over naively extending PPB to transient state. 

This paper is an extension of our previous work on rendering transient volumetric light transport~\cite{Marco2017transient}, where we proposed a naive extension of photon beams to transient state. Here we increase the applicability of the method, by proposing a progressive version of the space-time density estimation, and rigorously analyze its convergence.}

\section{Related Work}
Rendering participating media is a long-standing problem in computer graphics, with a vast literature on the topic. Here we focus on works related directly with the scope of the paper. For a wider overview on the field, we refer to the recent survey by Nov\'{a}k et al.~\cite{novak18monte}.

\paragraph*{Photon-based Light Transport.} 
Photon mapping~\cite{Jensen2001} is one of the most versatile and robust methods for rendering, with several extensions for making it suitable for animations~\cite{Cammarano2002}, adapting the distribution of photons~\cite{spencer2009into,Gruson2016spatial}, carefully selecting the radiance estimation kernel~\cite{spencer2009into,kaplanyan2013adaptive}, combining it with unbiased techniques~\cite{Georgiev2012,hachisuka2012path}, or making it progressive for ensuring consistency at limited memory requirements~\cite{hachisuka2008progressive,knaus2011progressive}. See \cite{Hachisuka2013course} for an in-depth overview. 
Jensen and Christensen~\cite{jensen1998efficient} extended photon mapping to media. Jarosz and colleagues significantly improved efficiency in volumetric photon mapping by introducing the beam radiance estimate \cite{Jarosz2008}. Generalization of beams to the tracing process by storing full photon trajectories (photon beams) \cite{Jarosz2011} led to a dramatic increase of density of photon maps at very little computational cost. Their progressive and hybrid counterparts \cite{jarosz2011progressive,krivanek14upbp} leveraged the benefits of beam radiance estimations while providing consistent solutions using finite memory. Recently, Bitterli and Jarosz~\cite{Bitterli2017beyond} proposed a generalization of photon beams to higher dimensions, proposing the use of photon planes, volumes and, in theory, higher-dimensional geometries, leading to unbiased density estimation. All these works are, however, restricted to steady-state renders; we instead focus on simulating light transport in transient state. 
\paragraph*{Transient rendering.} 
The transport equations~\cite{Chandrasekhar1960,Glassner1995principles} are time-resolved, most rendering algorithms focus on steady-state light transport. Still, several works have been proposed to deal with light transport in a time-resolved manner. In particular, most previous works on transient rendering have focused on simulating surfaces transport:  Klein et al.\cite{Klein2016SR} extended Smiths' transient radiosity~\cite{Smith2008} for second bounce diffuse illumination, while other works have used more general methods based on transient extensions of Monte Carlo (bidirectional) path tracing~\cite{Jarabo2012,Jarabo2014,Pitts2014,Jarabo2018bidirectional} and photon mapping~\cite{Meister2013,otoole2014}. Several works have also dealt with time-resolved transport on the field of neutron transport \cite{case1953introduction,bell1970nuclear,williams1971mathematical,duderstadt1979transport}.
Closer to our work, Ament and colleages~\cite{Ament2014} rendered transient light transport in refractive media using volumetric photon mapping. Jarabo et al.~\cite{Jarabo2014} proposed a transient extension of the path integral, and introduced an efficient technique for reconstructing the temporal signal based on density estimation. They also proposed a set of techniques for sampling media interactions uniformly in time. Finally, Bitterli~\cite{Bitterli2016} and Marco et al.~\cite{Marco2013,Marco2017transient} proposed a transient extension of the photon beams algoritm. 
Our work extends the latter, proposing a progressive, consistent, and robust method for rendering transient light transport. We leverage beams continuity and spatio-temporal density estimation to mitigate variance in the temporal domain, and derive the parameters for optimal convergence of the method.

\section{Transient Radiative Transfer}
\label{sec:transient_radiative_transfer}

%
The \emph{radiative transfer equation} (RTE) \cite{Chandrasekhar1960} models the behavior of light traveling through a medium. While the original formulation is time-resolved, its integral form used in traditional rendering ignores this temporal dependence, and computes the radiance $L$ reaching any point $\xm$ from direction $\omegav$ as
\begin{align} 
L(\xm, \omegav)\! &= \Tr(\xm, \xs)L_s(\xm_s, \omegav)+ \!\! \int_0^s\!\! \Tr (\xm, \xm_\mpar) L_o (\xm_\mpar, \omegav)\diff \mpar,
\label{eq:curveLFin}
\end{align}
where $\x_d=\xm-d\cdot\omegav$ is a point at distance $d$, $L_s$ is the radiance from the closest surface point $\xs$ at a distance $s$, $\Tr(\x,\x_t)=\exp(-\int_0^t \ext (\x_{t'})d\x_{t'})$ is the attenuation due to media between points $\x$ and $\x_t$ with $\ext(\x_t)$ the extinction at point $\x_t$, and $L_o$ is the in-scattered radiance at $\xm_t$ towards $\omegav$ 
\begin{align}
\changes{L_o} (\xm,\omegav) &= \sca(\xm)\int_{\Omega} \pf(\xm, \omegain, \omegav) \changes{L_i}(\xm,\omegain) \,\diff\omegain ,
\label{eq:outLF}
\end{align}
with $\Omega$ the sphere of directions, $\sca(\x_t)$ the scattering coefficient at point $\x_t$, $\pf$ the phase function, and $L_i(\xm,\omegain)$ the incoming radiance at point $\xm$ from direction $\omegain$. 

Equations \ref{eq:curveLFin} and \ref{eq:outLF} assume that the speed of light is infinite. However, if we want to solve the RTE at time scales comparable to the speed of light we need to incorporate the different delays affecting light. 
Light takes a certain amount of time to propagate through space, and therefore light transport from a point $\xm_0$ towards a point $\xm_1$ does not occur immediately, having (assuming light travels in straight lines)
\begin{align}
L(\xm_1, \omegav, t) = L(\xm_0, -\omegav, t - \Delta t),
\label{eq:timedependence}
\end{align}
where $\Delta t$ is the time it takes the light to go from $\xm_0$ to $\xm_1$. In turn, $\Delta t$ is defined by
\begin{align}
\Delta t (\xm_0 \leftrightarrow \xm_1)= \int_{\xm_0}^{\xm_1} \frac{\ior(\xm)}{c} \mathrm{d}\xm,
\label{eq:delta_t_integration}
\end{align}
where $\ior(\xm)$ is the index of refraction at a medium point $\xm$ and $c$ is the speed of light in vacuum. Note that in this case light does no travel in straight line, but by following the Eikonal equation~\cite{Ament2014,Gutierrez2005}. In a medium with a constant index of refraction $\ior(\xm) = \ior_m$, then $\Delta t (\xm_0 \leftrightarrow \xm_1)$ can be expressed as
\begin{align}
\Delta t (\xm_0 \leftrightarrow \xm_1) =\frac{\ior_m}{c} ||\xm_1 - \xm_0||.
\label{eq:delta_t_constant_ior}
\end{align}
The second form of delay occurs in the scattering events, and might occur from different sources, including electromagnetic phase shift, fluorescence and phosphorescence, or multiple scattering within the surface (or particle) microgeometry. To account for these sources of scattering delays, we introduce a temporal variable in the phase function as $\pf(\xm,\omegain,\omegav, t)$, where $t$ is the instant of light interacting with the particle before it is scattered.
%
%
With those delays in place, we reformulate the RTE (Equations \ref{eq:curveLFin} and \ref{eq:outLF}) introducing the temporal dependence as~\cite{Glassner1995principles}
\begin{align}
L(\xm,\omegav,t) &= T_r(\xm, \xs)L_s(\xm_s, \omegav, t-\Delta t_s)\eqbreak
&+ \int_{0}^{s}\! T_r(\xm, \xm_\mpar) L_o(\xm_\mpar, \omegav,t-\Delta t_\mpar) \mathrm{d}\mpar,
\label{eq:transient_rte_integral}
\end{align}
\begin{align}
\changes{L_o}\!(\xm, \omegav,t)\! =\sca(\xm)\int_{\Omega}\! \int_{-\infty}^t\!\!  \pf(\xm,\omegain,\omegav, t\!-\!t'\!) \changes{L_i}(\xm,\omegain,t) \diff t'\,\diff \omegain,
\label{eq:transient_inscattering}
\end{align}
with $\Delta t_s = \Delta t(\xm \leftrightarrow \xs)$ and $\Delta t_\mpar=\Delta t(\xm \leftrightarrow \xm_\mpar)$~\eqref{eq:delta_t_integration}. Note that we assume that the matter does not change at time-scales comparable to the speed of light, and therefore avoid any temporal dependence on $\sca$ and $\ext$. Introducing temporal variation at such speeds would produce visible relativistic effects~\cite{Weiskopf1999,Jarabo2015}.

\section{Transient Photon Beams}
\label{sec:tpb}
\begin{figure}[t]
	\centering
	\begin{subfigure}[t]{.32\columnwidth}
	\def\svgwidth{\columnwidth}  
	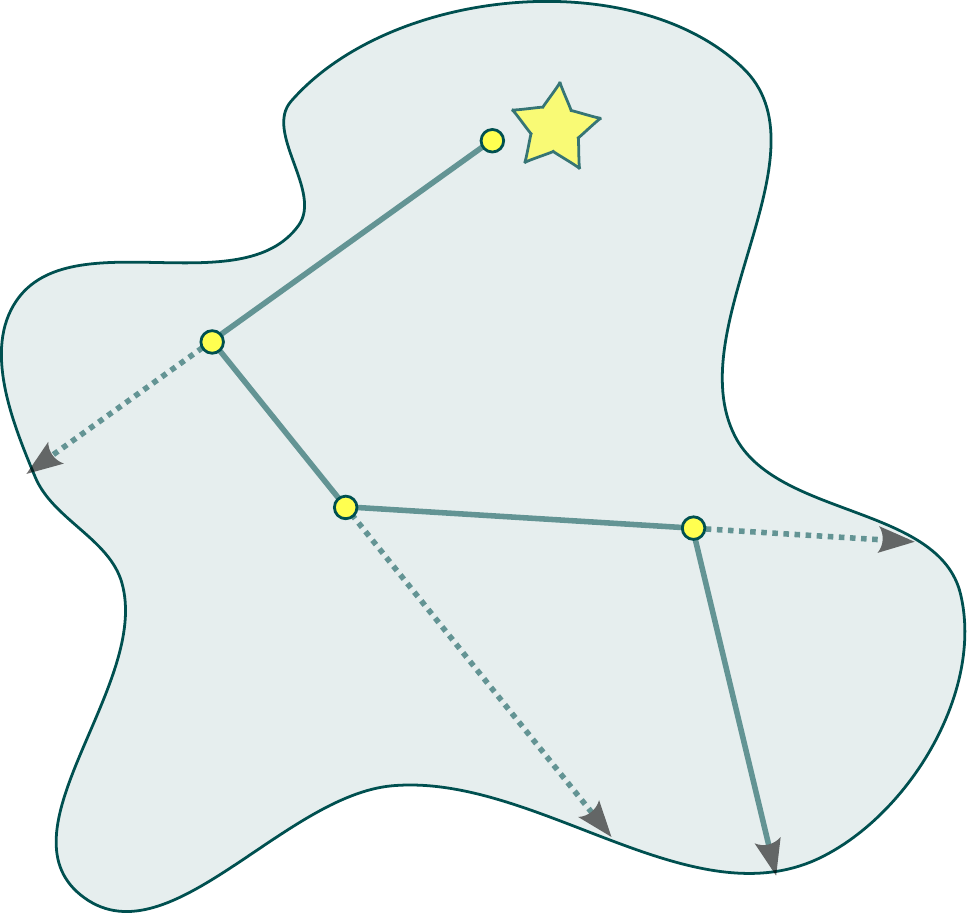
	\caption{} 
	\label{fig:photon_tracing_timing}
	\end{subfigure}
	\hspace{1em}
	\begin{subfigure}[t]{0.38\columnwidth}
	\centering
	\includegraphics[width=\textwidth]{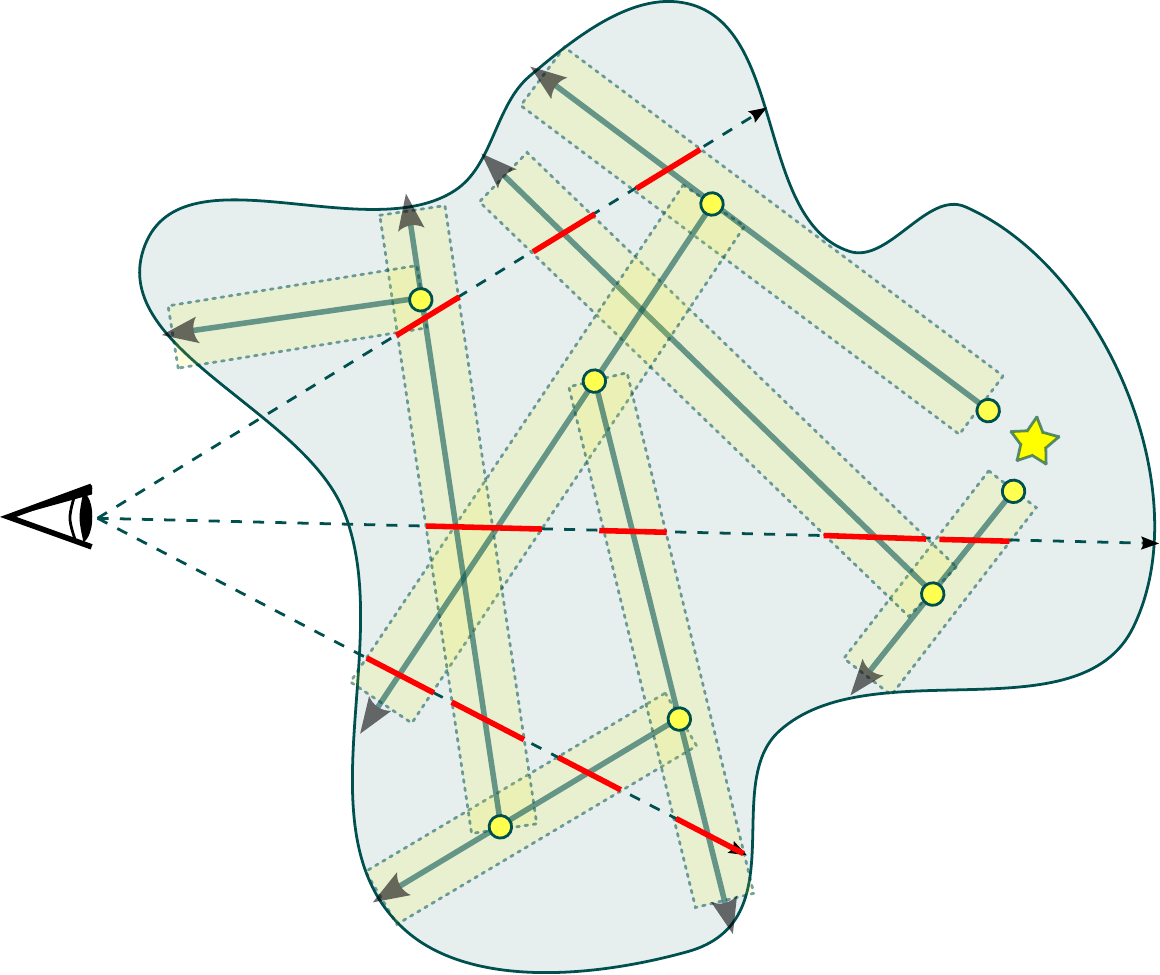}
	\caption{}
	\label{fig:beam_ray_full_estimation}
	\end{subfigure}
	\caption{(a) A photon emitted from the light source will take a time $t_{b_0} = \frac{\ior_m}{c} (s_1 + s_2 + s_3)$ to get $\xb$. (b) Radiance estimation in the medium is done by intersecting every ray against the photon beam map, and performing density estimations at the ray-beam intersections (red).}
\end{figure}

\begin{figure}[t]
	\centering
	\begin{subfigure}[b]{0.45\columnwidth}
		\def\svgwidth{\columnwidth}  
		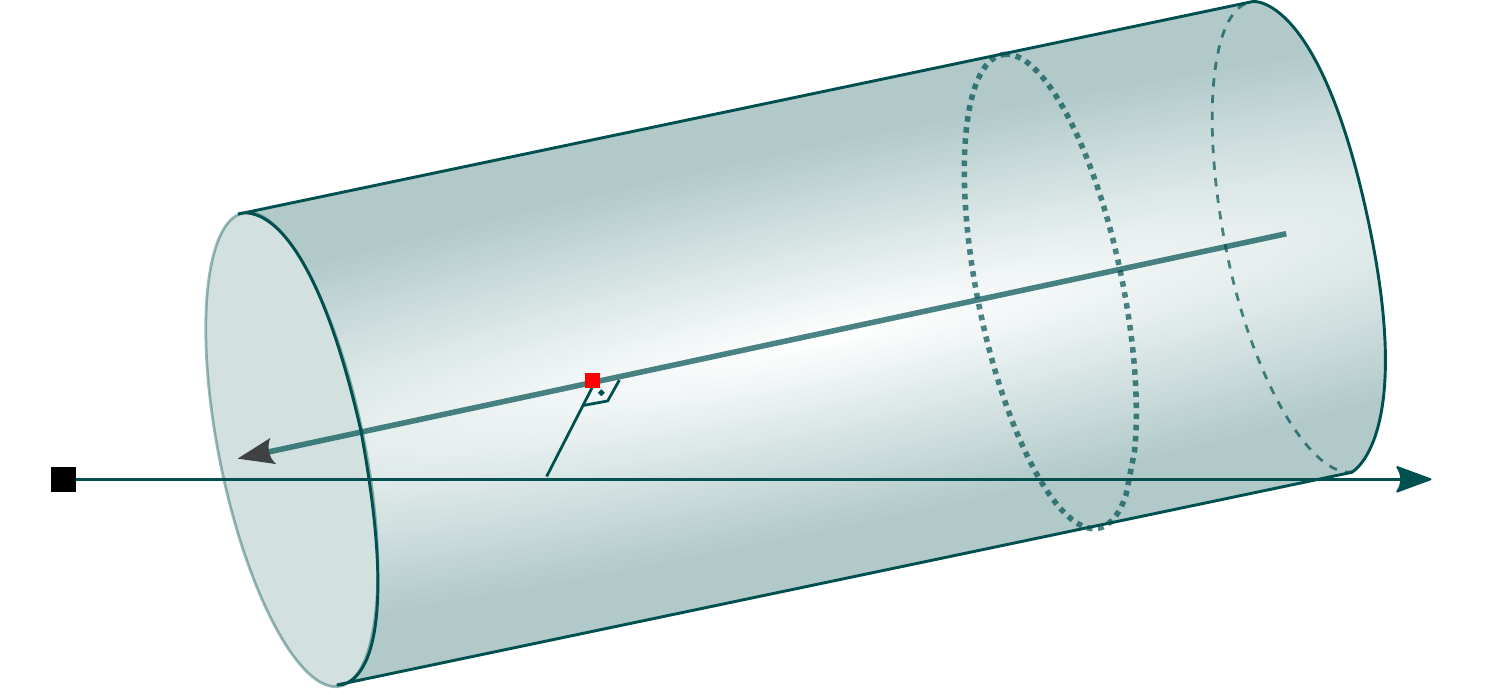
		\\\\
		\def\svgwidth{\columnwidth}  
		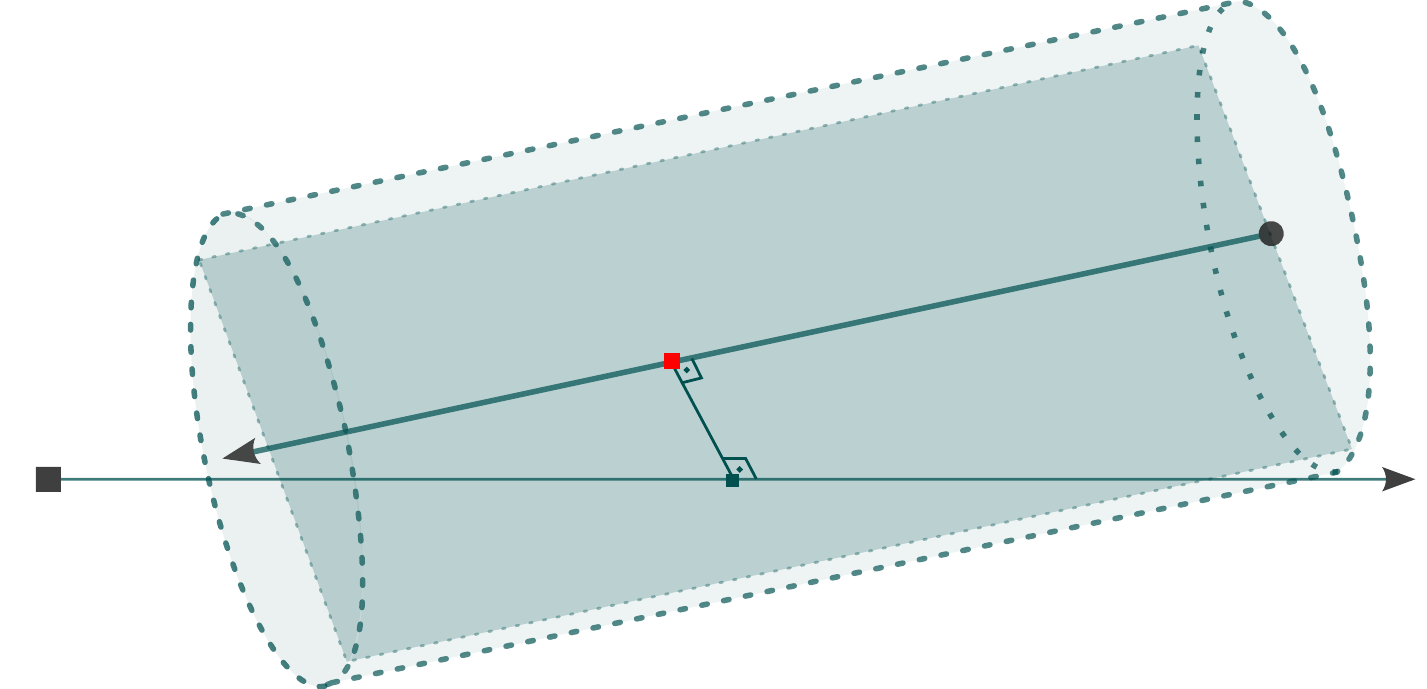
		\caption{} 
		\label{fig:rayq_beamd_2D_step3}
	\end{subfigure}
	\hspace{2em}
	\begin{subfigure}[b]{0.32\columnwidth}
		\def\svgwidth{\columnwidth}
		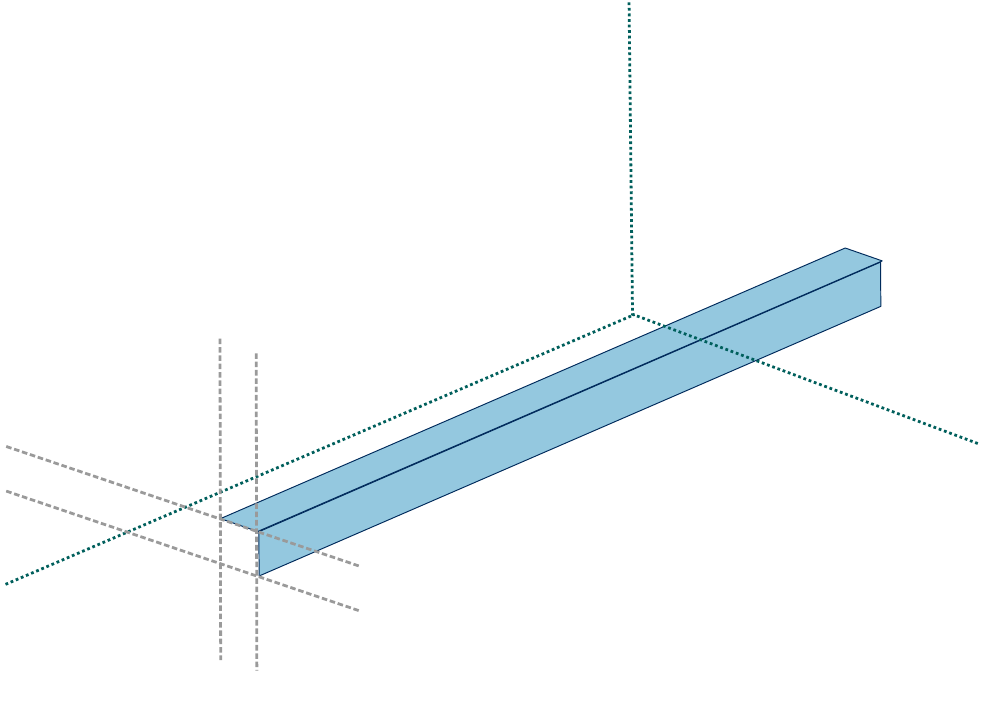
		\\
		\\
		\def\svgwidth{\columnwidth}
		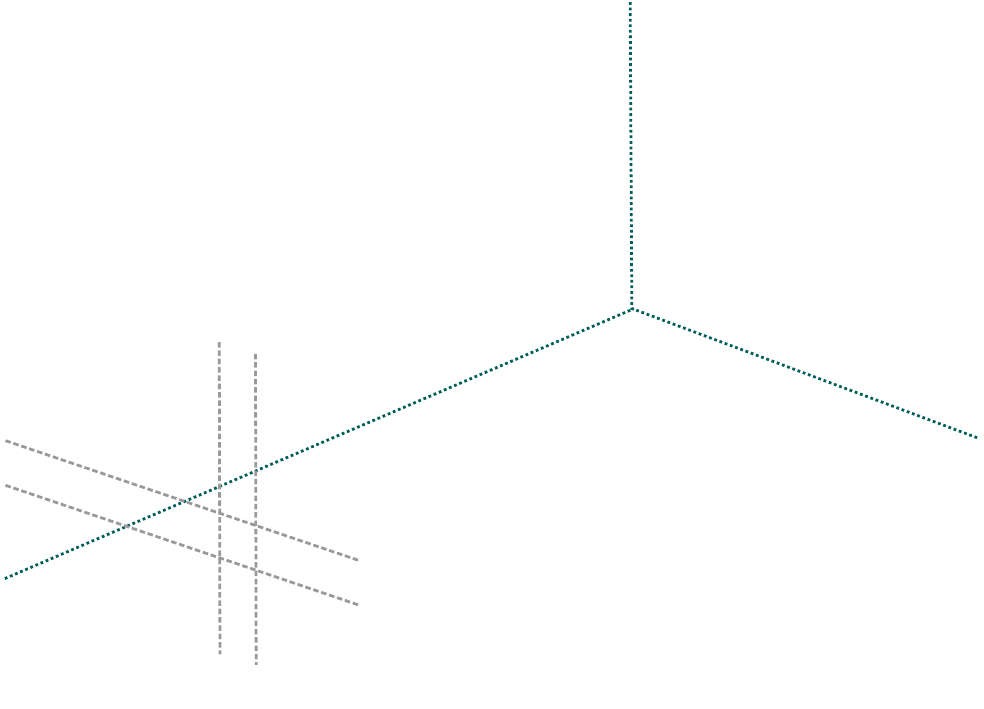
		\caption{} 
		\label{fig:volumetime}
	\end{subfigure}
	\caption{\forum{(a) Ray-beam intersection for density estimation using a 2D kernel (top) and 1D kernel (bottom). Time delays $t_b, t_r$ within these spatial density estimations will depend on the ray-beam orientation the blur region intersections ${s_b, s_r}$, the speed of light, and the index of refraction of the media. (b) Radiance estimate of a single beam at pixel $ij$ using a 2D blur generates a temporal footprint over a time interval $[t^-, t^+]$ (top) while radiance estimate using a 1D blur occurs at a single time instant $t$ (bottom).}}
\end{figure}
Photon beams \cite{Jarosz2011} provides a numerical solution for rendering participating media in steady state in two passes: In the first pass (Figure~\ref{fig:photon_tracing_timing}), a series of random walks are traced from the light sources. These paths represent packages of light (photons) traveling through the medium. Every interaction of a photon within the medium is stored on a map as a \emph{beam} with a direction $\omegav_b$, position $\xm_b$ and power $\ppwr_b$. 
%
In the second pass (Figure~\ref{fig:beam_ray_full_estimation}), rays are traced from the camera against the scene, and Equation~\eqref{eq:curveLFin} is approximated by summing up the contribution of all near photon beams $R_b$ of the eye ray defined by $r = (\xc, -\omegav_r)$
\begin{align}
L(\x_r, \omegav_r) \approx \sum_{b \in R_b} L_b(\xc, \omegav_r),
\label{eq:beam_ray_2D}
\end{align}
where $L_b(\xc, \omegav_r)$ is the contribution of photon beam $b$. 
Every photon beam $b$ is considered to have certain radius $R_b$, and radiance seen by a camera ray is computed by performing a density estimation on every ray-beam intersection. 

\subsection{Our algorithm} 
Our algorithm generalizes photon beams to transient state, so the same two steps are required. To move to transient state we need to introduce the temporal domain in the photon and eye random walks, which marks the temporal extent of photon beams, in form of both propagation and scattering delays, and also the effect of time in the paths \emph{merging} via density estimation. 

\paragraph*{Creating the photon map}
We compute the photon propagation as a standard random walk through the scene, which can be modeled using the subpath formulation defined by Jarabo et al.~\cite{Jarabo2014}. Let us define a light subpath $\bar{\xm}_l = \xm_0...\xm_k$, with $k$ vertices and $\xm_0$ at the light source. This light path defines $k-1$ photon beams, in which a beam $b_j$ is defined its origin at $\xm_{b_j}=\xm_j$ and direction $\omegav_{b_j}=\frac{\xm_{j+1}-\xm_j}{\|\xm_{j+1}-\xm_j\|}$. 
Using Jarabo's definition of the path integral (and therefore of the contribution of the subpaths), we compute the flux of each photon as:
\begin{equation}
\ppwr_{b_j} = \frac{f(\bar{\xm}_j,\bar{\sDeltaTime}_j)}{M p(\bar{\xm}_j,\bar{\sDeltaTime}_j)} = \frac{L_e(\xm_0\to\xm_1, \sDeltaTime_0) T(\bar{\xm}_j,\bar{\sDeltaTime}_j)}{M \prod_{i=0}^{j} p(\xm_i,\sDeltaTime_i)}, 
\end{equation}
with $\bar{\xm_j}$ the subpath of $\bar{\xm}_l$ up the vertex $j$, $\bar{\sDeltaTime_j} = \sDeltaTime_0...\sDeltaTime_j$ the sequence of time delays up to vertex $j$, $M$ the number of photon random walks sampled, $L_e(\xm_0\to\xm_1, \sDeltaTime_0)$ the emission function, $p(\xm_i,\sDeltaTime_i)$ the probability of sampling vertex $\xm_i$ with time delay $\sDeltaTime_i$. $T(\bar{\xm_j},\bar{\sDeltaTime_j})$ the throughput of subpath $(\xm_i,\sDeltaTime_j)$ defined as:
 \begin{equation}
T(\bar{\xm}_j,\bar{\sDeltaTime}_j) = \left[\prod_{i=1}^{j-1} \pf(\xm_i,\sDeltaTime_j) \right] \left[\prod_{i=0}^{j-1} G(\xm_i,\xm_{i+1}) V(\xm_i,\xm_{i+1})\right],
\end{equation}
with $\pf(\xm_i,\sDeltaTime_j)$ the scattering event at vertex $\xm_i$ with delay $\sDeltaTime_j$, and $G(\xm_i,\xm_{i+1})$ and $V(\xm_i,\xm_{i+1})$ the geometry and visibility terms between vertices $\xm_i$ and $\xm_{i+1}$, respectively. 
Finally, for transient state we need to know the instant $t_{b_j}$ at which the photon beam is created (through emission or scattering), defined as:
\begin{equation}
t_{b_j} = \sum_{i=0}^{j-1} \sDeltaTime_j + \sum_{i=0}^{j-1} \Delta t(\xm_i,\xm_{i+1}). 
\end{equation}

\paragraph*{Rendering}
For rendering, we adapt Equation~\eqref{eq:beam_ray_2D} to account for the temporal domain, as
\begin{align}
L(\xc, \omegav_r,t) \approx \sum_{b \in R_b} L_b(\xc, \omegav_r,t),
\label{eq:beam_ray_2D_transient}
\end{align}
with $L_b(\xc,\omegav_r,t)$ the radiance estimation for beam $b$ to ray $t$ at instant $t$, with $b(s_b) = \xb+s_b\cdot\omegav_b$  and  $r(s_r) = \xc - s_r\cdot\omegav_r$. In essence, $L_b(\xc,\omegav_r,t)$ will return zero radiance if $t$ is out of the temporal footprint of the density estimation kernel. Depending on the dimensionality of the density estimation, Jarosz and colleagues~\cite{Jarosz2011} proposed three different estimators based on 3D, 2D and 1D kernels. \forum{Since the 3D kernel results impractical due to costly 3D convolutions, we focus on 1D and 2D kernels. In the following we extend Jarosz et al.'s 2D and 1D kernels to transient state, assuming homogeneous media. }

\paragraph*{Kernel 2D} We generalize Jarosz's et al.'s 2D estimate $L_{b|\textrm{2D}}$ by introducing a temporal function $W(t)$ as
\begin{align}
L_{b|\textrm{2D}}(\xc, \omega_r,t) 
\label{eq:single_beam_ray_2D_closedform}
= & K_{\textrm{2D}}(R_b)\ppwr_b \pf(\theta_b) \sca \nonumber \\ & \frac{e^{-\ext(s_c^- - s_c^+)(\mid\cos\theta_b\mid-1)} - 1}{e^{\,\ext (s_r^- + s_b^-)} \ext (|\cos\theta_b| - 1)} W_{2D}(t),
\end{align}
where $[s_r^-,s_r^+]$ are the limits of the ray-beam intersection (\fref{fig:rayq_beamd_2D_step3}), $\theta_b$ is the angle between $\omegav_b$ and $\omegav_r$, and $K_{\textrm{2D}}(R_b)$ is a canonical 2D kernel with radius $R_b$. The temporal function $W_{2D}(t)$ models the temporal footprint of the 2D kernel as
\begin{equation}
W_{2D}(t) = 
  \begin{cases}
    \frac{1}{t^+-t^-}  & \quad \text{if } t \in (t^-,t^+)   \\
    0  & \quad \text{otherwise}
  \end{cases},
\end{equation}
where $t^- = t_b + t_r + \frac{\ior_m}{c} (s_r^-+s_b^-)$ and $t^+ = t_b + t_r + \frac{\ior_m}{c} (s_r^++s_b^+)$, and $t_r$ the initial time of the camera ray, computed similarly to $t_b$.  
Note that due to transmittance, the photon energy varies as it travels across the blur region. Evenly distributing the integrated radiance $L_b$ across this interval introduces temporal bias, in addition to the inherent spatial bias introduced by density estimation. 
However we observed this even distribution provides a good tradeoff between bias, variance, and computational overhead.

\paragraph*{Kernel 1D} 
\forum{In the 1D kernel defined for density estimation by Jarosz et al. the spatial blur is performed over a line. Therefore, the energy of the beam is just spread on the ray on a single point at $r(s_r)$, from a single point of the beam $b(s_b)$ (see Figure~\ref{fig:rayq_beamd_2D_step3}). In consequence, $s_r^\pm\to s_r$ and $s_b^\pm\to s_b$, which implies that $t^\pm\to t_{br}$, and the temporal function reduces to $W_{1D}(t-tb)=\delta(t)$, with $\delta(t)$ the Dirac delta function. 
With that in place, we transform Jarosz et al. 1D estimate to
\begin{align}
L_{b|\textrm{1D}}(\xc, \omega_c,t) 
\label{eq:single_beam_ray_1D_closedform}
&= K_{\textrm{1D}}(R_b)\ppwr_b \pf(\theta_b) \sca \frac{e^{-\ext s_b} e^{-\ext s_r}}{\sin\theta_b}\delta(t-t_b),
\end{align}
with $K_{\textrm{1D}}(R_b)$ a 1D kernel with radius $R_b$. }

\paragraph*{Implementation} 

\forum{Since photon beams correspond to full photon trajectories, they allows us to estimate radiance at any position $\xb+s\omegav_b$ of the beam, and therefore at any arbitrary time $t(\xb+s\omegav_b)$. As mentioned, one-dimensional radiance estimate corresponds to a single time across the beam. In a traditional rendering process where camera rays are traced through view-plane pixels against the beams map, the temporal definition \textit{within} a pixel will be proportional to the amount of jittering performed at the pixel level. Additionally, 2D blur requires distributing every radiance estimate along a time interval, which reduces variance in the time dimension of a pixel at the expense of introducing additional temporal bias.}

\forum{Finally, note that the temporal footprint of the density estimation might be arbitrarily small, so the probability of finding a beam $b$ at an specific time might be very low. We alleviate this issue using path reuse via density estimation~\cite{Jarabo2014}. In particular, for the non-progressive results we use the histogram density estimation. In Section~\ref{sec:ptpb} we introduce temporal kernel-based density estimation, and combine it with the spatial density estimation of the beam.}


\section{Progressive Transient Photon Beams}
\label{sec:ptpb}

\forum{%
Kernel density estimation reduces variance at the expense of introducing bias in the results, which makes both Equations~\eqref{eq:beam_ray_2D} and ~\eqref{eq:beam_ray_2D_transient} to not converge to the actual solution, even with an infinite number of photons $M$. 
In order to avoid this undesirable convergence, progressive density estimation aims to provide a biased, yet consistent technique, that in the limit converges to the expected value (in other works, the bias vanishes in the limit). The key idea is to average several render passes with a finite number of photon random walks $M$, progressively reducing the bias in each iteration while allowing variance to slightly increase. 
In order to fully leverage progressive density estimation, we extend the spatial density estimation in Section~\ref{sec:tpb} to the temporal domain. In the following, we present our spatio-temporal beam density estimation, and then present our progressive approach.

\paragraph*{Spatio-Temporal Beam Estimation}
Jarabo et al.~\cite{Jarabo2014} shown that progressive density estimation in the temporal domain can in fact increase the convergence for transient renderer, in particular when compared with the histogram method used in Section~\ref{sec:tpb} for rendering the temporal domain. To combine such approach with the (progressive) spatial density estimation in photon beams~\cite{jarosz2011progressive}, we reformulate the 1D kernel in Equation~\eqref{eq:single_beam_ray_1D_closedform}, by convolving it with a 1D temporal kernel $\sKt(t)$ so that
%
\begin{align}
L_{b|\textrm{1D}}(\xc, \omega_c,t) 
	&= \sKoneD(R_b)\ppwr_b \pf(\theta_b) \sca \frac{e^{-\ext s_b} e^{-\ext s_r}}{\sin\theta_b}K_\sT(t-t_b).
\label{eq:single_beam_ray_1D_closedform}
\end{align}

\paragraph*{Progressive Transient Photon Beams}
We generalize the computation of $L(\xc, \omegav_r,t)$~\eqref{eq:beam_ray_2D_transient} using an iterative estimator, defined as
\begin{align}
L(\xc, \omegav_r,t) \approx \estimate{\sPixel}_n(\xc, \omegav_r,t) = \frac{1}{n}\sum_{i=0}^n \sum_{b \in B_i} L_b(\xc, \omegav_r,t)
\label{eq:estimate}
\end{align}
with $\estimate{\sPixel}_n$ the estimate of $L$ at $n$ iterations, and $B_i$ the set of photon beams per iteration $i$. Note that the previous equation assumes that the camera ray $r$ is the same for all iterations. That is not necessarily true (and in fact it is not) but for simplicity we express this way. 

The error of the estimate $\estimate{\sPixel}_n$ is defined by its bias and variance, which as shown in Appendix~\ref{sec:error_beam_radiance_estimate} is dependent on the bandwidth of the spatial and temporal kernels. In particular, the variance of the error increases linearly with the bandwidth of the kernels, while bias is reduced at the same rate. 
Then, on each iteration we reduce the bias by allowing the variance to increase at a controlled rate of $(i+1)/(i+\alpha)$, with $\alpha\in[0,1]$ being a parameter that controls how much the variance is allowed to increase at each iteration.To achieve that reduction, on each iteration $i+1$ we reduce the footprint of kernels $K_{\textrm{1D}}$ and $K_\sT$ ($\sR_{|j}$ and $\sT_j$) by
\begin{align}
\frac{\sR_{|j+1}}{\sR_{|j}} = \left(\frac{j+\alpha}{j+1}\right)^{\beta_R} &,  &
\frac{\sT_{j+1}}{\sT_j} = \left( \frac{j+\alpha}{j+1} \right)^{\beta_\sT} &,
\label{eq:tppm_alpha}
\end{align}
where $\beta_R$ and $\beta_\sT$ control the individual reduction ratio of each kernel, with $\beta_\sT=1-\beta_R$. In the following, we analyze the convergence rate of the method, and compute the optimal values for the parameters $\alpha$, $\beta_\sT$ and $\beta_R$. 

\paragraph*{Convergence analysis} 
We analyze the convergence of the algorithm as a function of the \textit{asymptotic mean squared error} (AMSE) defined as
\begin{equation}
\AMSE(\estimate{\sPixel}_n) = 
	\sVar{\estimate{\sPixel}_n} +
	\sEV{\sError_n}^2,
	\label{eq:amse}
\end{equation}
where $\sVar{\estimate{\sPixel}_n}$ is the variance of the estimate and $\sEV{\sError_n}$ is the bias at iteration $n$. As shown in Appendix~\ref{sec:app_convergence}, the variance converges with rate 
\begin{eqnarray}
\sVar{\estimate{\sPixel}_n} \approx
	\Order{n^{-1}} + \Order{n^{-\alpha}} = \Order{n^{-\alpha}},
\label{eq:asymp_var}
\end{eqnarray}
while the bias converges with rate
\begin{equation}
\sEV{\sError_n} =
	\Order{n^{1-\alpha}}^{-2\scaleParT}+\Order{n^{1-\alpha}}^{2 \scaleParT-2}.
\label{eq:asymp_bias}
\end{equation}

Plugging Equation~\eqref{eq:asymp_var} and~\eqref{eq:asymp_bias} into Equation~\eqref{eq:amse}, we can model the AMSE as
\begin{eqnarray}
\AMSE(\estimate{\sPixel}_n) =
	\Order{n^{-\alpha}} + \left(\Order{n^{1-\alpha}}^{-2\scaleParT} +
	\Order{n^{1-\alpha}}^{2 \scaleParT-2}\right)^2.
	\label{eq:tppm_amse_text}
\end{eqnarray}

Finally, by minimizing Equation~\eqref{eq:tppm_amse_text} (see Appendix~\ref{sec:minimizing_amse}) we obtain the values for optimal asymptotic convergence $\scaleParT = 1/2$ and $\alpha = 2/3$, which by substitution gives us the final asymptotic convergence rate of our progressive transient photon beams 
\begin{equation}
\AMSE(\estimate{\sPixel}_n) = \Order{n^{-\frac{2}{3}}}.
\end{equation}%
}%

\section{Results}
\label{sec:results}

\begin{figure}[!t]
	\centering
	\includegraphics[width=0.8\columnwidth]{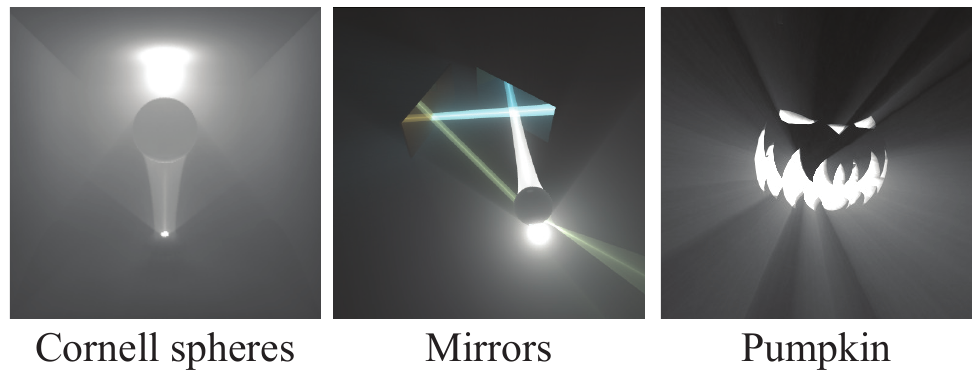}
	\caption{Steady-state renders for the scenes \textsc{Cornell spheres} (\fref{fig:cornell_spheres}), \textsc{Mirrors} (\fref{fig:mirrors}), and \textsc{Pumpkin} (\fref{fig:pumpkin_convergence}).}
	\label{fig:steady_state}
\end{figure}

\begin{figure}[!t]
	\centering
	\includegraphics[width=0.95\columnwidth]{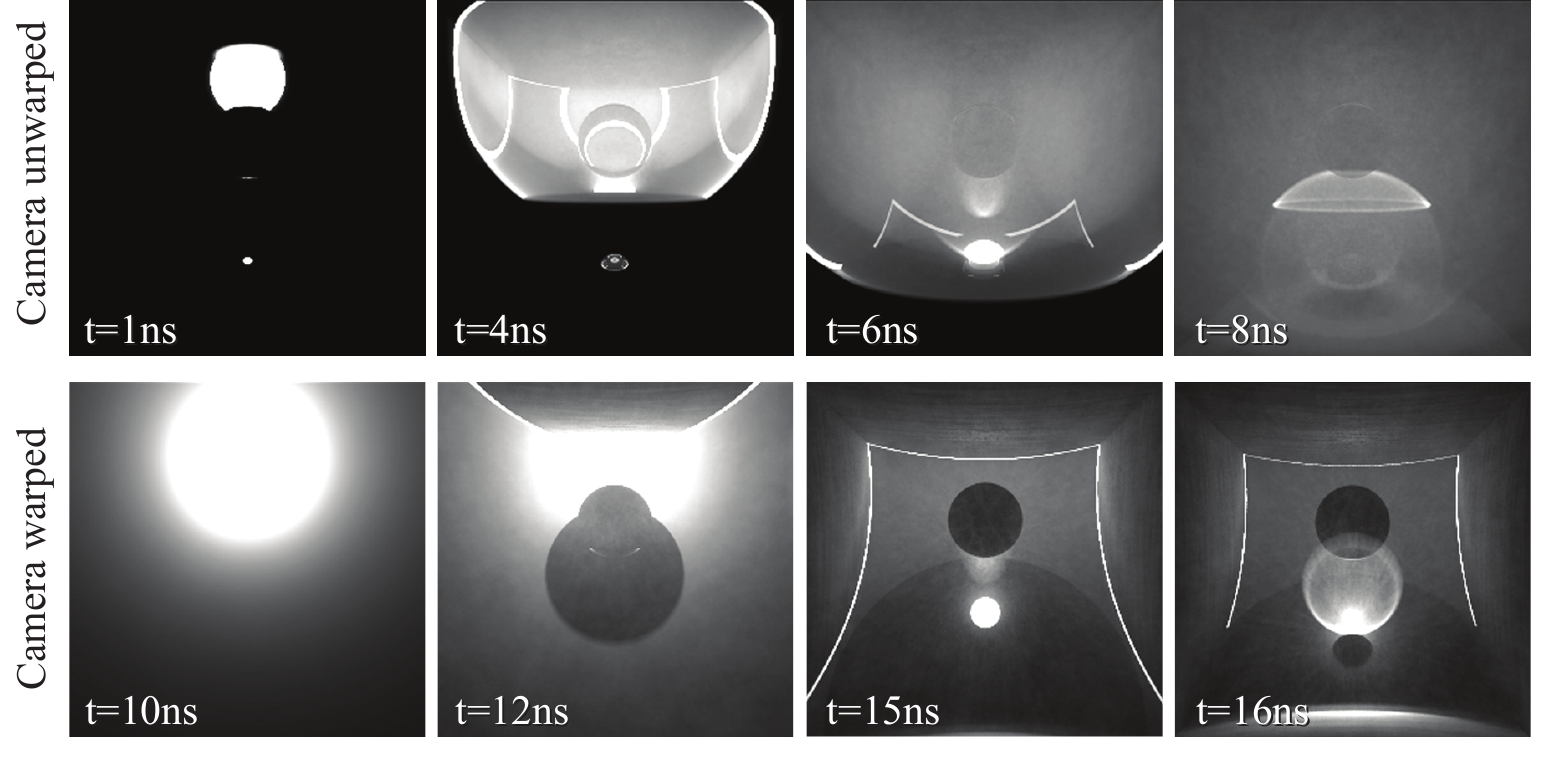}
	\caption{Comparison of \textsc{Cornell spheres} scene using \emph{camera-unwarping} (top), where we do not take into account the camera time, and real propagation of light (bottom). In the bottom row the shape of the wavefront is altered by the camera time, as if we were scanning the scene from the viewpoint towards the furthest parts of the scene. Camera unwarping on the other hand illustrates more intuitively how light propagates locally.}
	\label{fig:cornell_spheres}
\end{figure}

\begin{figure*}[!h]
	\centering
	\includegraphics[width=\textwidth]{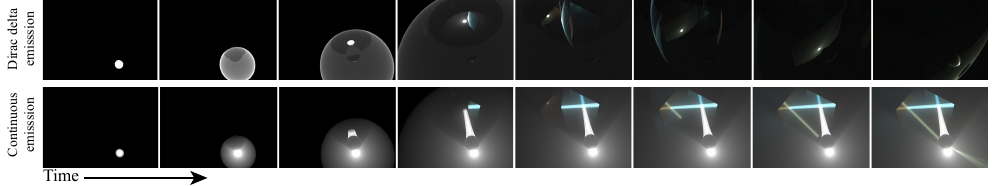}\vspace{-0.2em}
	\caption{Comparison between Dirac delta (top) and continuous (Heaviside) emission (bottom).  Dirac delta emission lets us see how a pulse of light travels and scatters across the scene, depicting the light wavefronts bouncing on the mirrors and going through the glass ball. Continuous emission shows how light is emitted until it reaches every point in the scene, as if we were taking a picture with a camera at very slow-motion.}
	\label{fig:mirrors}
\end{figure*}

\begin{figure*}[!h]
	\centering
	\includegraphics[width=\textwidth]{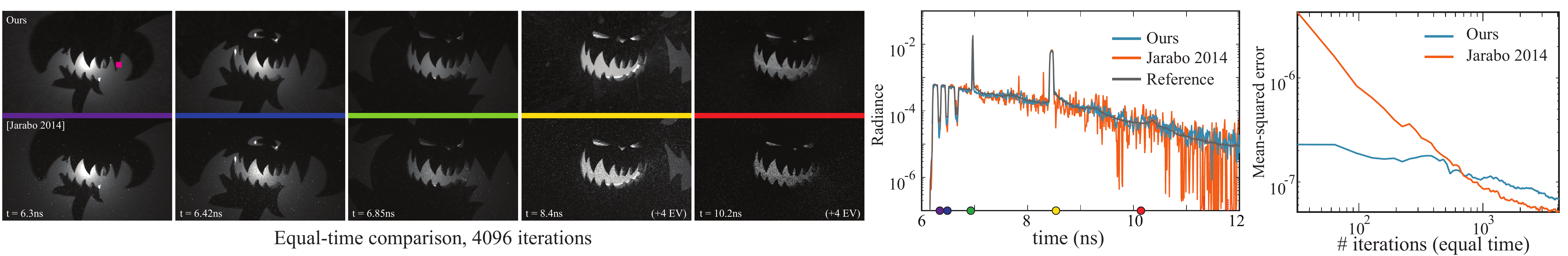}
	\caption{\forum{The \textsc{Pumpkin} scene shows a jack o'lantern embedding a point light that creates hard shadows through the holes. The left frames show a sequence of the time-resolved renders after 4096 iterations of our algorithm (10k beams / iteration), and temporal KDE on a progressive transient path tracer (PTPT, 16spp / iteration) \cite{Jarabo2014}. The middle plot compares the whole temporal footprint at the pink marker. Reference solution (dark grey) was obtained with a transient path tracer (no KDE) using 64M samples per pixel. Right plot shows MSE convergence with respect to the number of progressive iterations (in log-log scale), at 1 minute/iteration on each algorithm. As expected, the convergence of our method ($\Order{n^{-\frac{2}{3}}}$) is slower than PTPT ($\Order{n^{-\frac{4}{5}}}$) ; however, as shown in the equal-time comparison, our algorithm presents better temporal behavior with much less variance on later timings.}}
	\label{fig:pumpkin_convergence}
\end{figure*}

\begin{figure*}[!h]
	\centering
	\includegraphics[width=\textwidth]{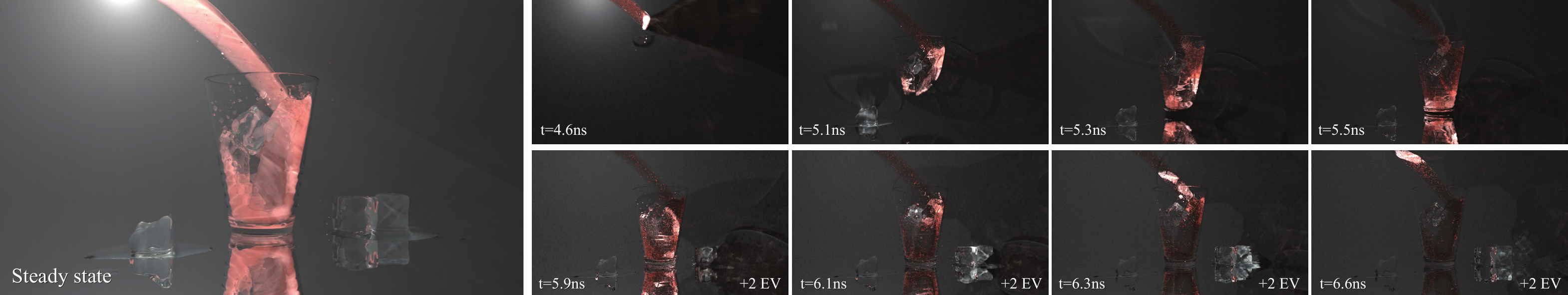}
	\caption{\forum{We illustrate the potential of our method in the \textsc{Juice} scene~\cite{bitterli16resources}, which presents a scene very difficult to render for path tracing methods, but well-handled by photon-based methods. The scene is filled by a thin participating medium, while the glass contains ruby grapefruit juice as measured by Narasimhan et al. \cite{narasimhan2006acquiring}. The highly forward phase function of the juice, as well as the delta interactions on the glass, ice cubes, and the mirror floor surface, generate complex caustic patterns which our method is able to simulate in transient state. Bottom row has increased exposure respect to top row to show the radiance at later timings.}}
	\label{fig:juice}
\end{figure*}

In the following we illustrate the results of our proposed method in five scenes: \textsc{Cornell spheres}, \textsc{Mirrors}, \textsc{Pumpkin}, \textsc{Soccer} \cite{sun2010line}, \textsc{Pumpkin}, and \textsc{Juice}. See Figures~\ref{fig:steady_state}, \ref{fig:soccer} (right), and \ref{fig:juice} (left) for steady-state renders of the scenes. 
Results of Figures \ref{fig:cornell_spheres} and \ref{fig:mirrors} were taken on a desktop PC with Intel i7 and 4GB RAM using a transient 2D kernel (Equation \ref{eq:single_beam_ray_2D_closedform}). 
\forum{Figures \ref{fig:soccer}, \ref{fig:pumpkin_convergence}, and \ref{fig:juice} were rendered on an Intel Xeon E5 with 256GB RAM, using our progressive spatio-temporal kernel density estimations (Section \ref{sec:ptpb}) derived from the transient spatial 1D kernel (Equation \ref{eq:single_beam_ray_1D_closedform}). 
All temporal density estimations are performed using radiance samples within fixed radius of the corresponding iteration (instead of using a nearest neighbor approach). Please refer to the supplemental video for the full sequences of all the scenes.}
%




\fref{fig:cornell_spheres} shows a Cornell box filled with a scattering medium, and demonstrates the effect of \emph{camera unwarping} \cite{Velten2013} when rendering. Camera unwarping is an intuitive way of visualizing how light propagates \emph{locally} on the scene without accounting for the time light takes to reach the camera. The scene consists of a diffuse Cornell box with a point light on the top, a glass refractive sphere (top, IOR = 1.5) and a mirror sphere (bottom). While \fref{fig:cornell_spheres}b shows the real propagation of light---including camera time---, \fref{fig:cornell_spheres}a depicts more intuitively how light comes out from the point light, travels through the refractive sphere, and the generated caustic bounces on the mirror sphere. Note how in the top sequence we can clearly see how light is slowed down through the glass sphere due to the higher index of refraction. We can also observe multiple scattered light (particularly noticeable in frames t=4ns and t=6ns) as a secondary wavefront.

\fref{fig:mirrors} compares visualizations of light propagation within the \textsc{Mirrors} scene using Heaviside and Dirac delta light emission. The scene is composed by two colored mirrors and a glass sphere with IOR = 1.5, and was rendered using the previously mentioned camera unwarping. We can observe how delta emission generates wavefronts that go through the ball and bounce in the mirrors, creating wavefront holes where constant emission creates medium shadows. In the last frame of the top row Delta emission clearly depicts the slowed down caustic through the glass ball respect to the main wavefront.

\forum{
Our progressive method combines time-resolved 1D spatial kernels of photon beams and temporal density estimations, reducing bias while providing consistent solutions in the limit with an optimal convergence rate of $\Order{n^{-\frac{2}{3}}}$. In \fref{fig:pumpkin_convergence} we analyze its convergence with respect to progressive transient path tracing with temporal KDE \cite{Jarabo2014} (PTPT). In the middle graph we show the temporal profile on a single pixel for both our algorithm and PTPT after 4096 equal-time iterations, where both algorithms converge to the reference solution taken with transient path tracing (no temporal KDE) with 64 million samples. While PTPT presents faster convergence (see \fref{fig:pumpkin_convergence}, right graph), our algorithm presents a better behavior over time where variance increases due to the lack of samples (center graph). Additionally, it requires much fewer iterations than PTPT to achieve a similar MSE (see log-log right graph).

In \fref{fig:soccer} we show a more complex scenario, with different caustics rendered, with our progressive algorithm. It contains a smooth dielectric figurine with different transmission albedos placed within a participating medium with an isotropic phase function. Our method is capable of handling complex caustics transmitted from light sources through the player, and then through the ball. Our algorithm progressively reduces bias and variance to provide a consistent solution.
}

\forum{
Finally in \fref{fig:juice} we illustrate a setup combining different media properties, and specular refractive and reflective materials. The liquid has a very forward phase function, making the light first travel through the direction of the stream ($t=4.6$ ns), and then going through the liquid inside the glass ($t=5.1$ns to $t=6.3$ns). The mirror surface makes the light to bounce back to the surrounding medium as a caustic through the water spills and ice cubes at $t=5.1$ns and $t=6.6$ns. Note that these are not fully observable in the steady-state render (left) due to the accumulated radiance from the surrounding medium and the adjusted exposure of the image. }
\vspace{-0.0em}
\section{Conclusions}
In this paper we have presented a robust progressive method for efficiently rendering transient light transport with consistent results. We derived our method based on progressive photon beams \cite{jarosz2011progressive}, extending its density estimators to account for light time-of-flight, and \forum{deriving a new progressive scheme. We then compute the convergence of the method, and derive the parameters for optimal asymptotic convergence. Our results demonstrate that combining continuous photon trajectories in transient state and our optimal spatio-temporal convergence rates allow to robustly compute a noise-free solutions to the time-resolved RTE for complex light paths.}
%
We believe that out work might be very useful for developing new techniques for transient imaging and reconstruction in media, as well as to obtain new insights on time-resolved light transport. 
%

As future work it would be interesting to analyze more thoroughly the optimal performance and kernels for variance reduction and bias impact in transient state, under varying media characteristics. In addition, extending our method to leverage recent advances in media transport, such as transient-state adaptations of higher-dimensional photon estimators \cite{Bitterli2017beyond} as well as hybrid techniques \cite{krivanek14upbp}, could improve performance of time-resolved rendering for a general set of geometries and media characteristics.

\appendix
\forum{
\section{Error in Transient Progressive Photon Beams}

Here we analyze the consistency of the transient progressive photon beams algorithm described in Section~\ref{sec:ptpb}.
For our analysis on the error of the estimate, we use the \textit{asymptotic mean squared error} (AMSE) defined as
\begin{equation}
\AMSE(\estimate{\sPixel}_n) = 
	\sVar{\estimate{\sPixel}_n} +
	\sEV{\sError_n}^2,
\end{equation}%
where $\sVar{\estimate{\sPixel}_n}$ is the variance of the estimate and $\sEV{\sError_n}$ is the bias at iteration $n$. 
We model $\sVar{\estimate{\sPixel}_n}$ as~\cite{knaus2011progressive}%
\begin{equation}
\sVar{\estimate{\sPixel}_n}  = 
	\frac{1}{n} \sVar{\sEyeContribution \ \Lo} +
	\frac{1}{n^2} \sum_{j=1}^n\sVar{\sEyeContribution \ \sError_j},
\label{eq:app_variance_n}
\end{equation}
where $\sEyeContribution$ is the contribution of the eye ray, and $\sError_j$ is the bias for iteration $j$. The first term is the standard variance of the Monte Carlo estimate, which is unaffected by the kernel. The second term, on the other hand, is the variance of the error, and is dependent on density estimation. 
On the other hand, the estimated value of the error (bias) \sEV{\estimate{\sPixel}_n} is defined as
\begin{eqnarray}
\sEV{\estimate{\sPixel}_n} = \sPixel + \sEV{\sEyeContribution}\sEV{\sError_n},
\label{eq:tppm_ev_pixel_estimate}
\end{eqnarray}%
where $\sEV{\sError_n}$ is the bias of the estimator after $n$ steps:%
\begin{equation}
	\sEV{\sError_n} = \frac{1}{n} \sum_{j=1}^n \sEV{\sError_j},
	\label{eq:tde_bias_n}
\end{equation}%
with $\sEV{\sError_j}$ the expected error at iteration $j$. 
In the following, we first derive the variance and expected value of the error for a single iteration. Then, we analyze the asymptotic behavior of the these terms, and compute the values for optimal convergence for $\beta_\sT$, $\beta_R$ and $\alpha$.  
\section{Variance and Expected Value of the Error of the Time-Resolved Beam Radiance Estimate}
\label{sec:error_beam_radiance_estimate}
We first analyze the variance and expected value of the error (bias) introduced by the radiance estimate at each iteration. Let us first define the error in each iteration as:
\begin{align}
\sError & =\estimate{\sPixel}_n(\xc, \omegav_r,t) - L(\xc, \omegav_r,t) \nonumber \\ 
&= \sum_{i=1}^M\sKoneD(R_b) \sKt(t-t_i) \sLightContribution_i  - \Lo(\xc, \omegav_r,t).
\label{eq:error}
\end{align}

\paragraph{Variance} We first define the variance of the error $\sVar{\sError}$ as (in the following, we omit dependences for clarity):
\begin{eqnarray}
\sVar{\sError} & = & \sVar{\sum_{i=1}^M\sKoneD \sKt \sLightContribution  - \Lo} \\
		  & = & (\sVar{\sKoneD}+\sEV{\sKoneD}^2)(\sVar{\sKt}+\sEV{\sKt}^2) \nonumber \\ 
& & (\sVar{\sLightContribution}+\sEV{\sLightContribution}^2)-\sEV{\sKoneD}^2\sEV{\sKt}^2\sEV{\sLightContribution}^2, \nonumber
\label{eq:tppm_var_err}
\end{eqnarray}
In order to compute the variance of the error $\sVar{\sError}$ we need to make a set of assumptions: First, we assume that the beams' probability density is constant within the kernel $\sKoneD$ in the spatial domain~\cite{jarosz2011progressive}, and within $\sKt$ in the temporal domain~\cite{Jarabo2014}. We denote these probabilities as $\sProbabilityR$ and $\sProbabilityT$ respectively. We also assume that the distance between view ray and photon beam, time $\sTime_b$ and beams' energy $\sLightContribution_i$ are independent samples of the random variables $\sRandomVariableR$, $\sRandomVariableT$ and $\ppwr$, respectively, which are mutually independent. Finally, we assume that $\sRandomVariableR$ and $\sRandomVariableT$ have probability densities $\sProbabilityR$ and $\sProbabilityT$. 

With these assumptions, and taking into account that $\sEV{\sKoneD}=\sProbabilityR$ and $\sEV{\sKt}=\sProbabilityT$, we can model the the variance introduced by the temporal kernel $\sVar{\sKt}$ as~\cite{Jarabo2014}%
\begin{equation}
\sVar{\sKt} = 
	\frac{\sProbabilityT}{\sT}
	\int_{\Real}
	\sKtCanonical(\sKVarChange)^2\diff\sKVarChange - \sProbabilityT^2, 
\end{equation}
where we express $\sKt$ as a canonical kernel $\sKtCanonical$ with unit integral such that $\sKt(\sKVar)=\sKtCanonical(\sKVar/\sT)\sT^{-1}$. Analogously, $\sVar{\sKoneD}$ is \cite{jarosz2011progressive}:%
\begin{equation}
\sVar{\sKoneD} =
	\frac{\sProbabilityR}{\sR}
	\int_{\Real}\sKoneDCanonical(\sKVarChange)^2\diff\sKVarChange -
	\sProbabilityR^2 .
\end{equation}%
This allow us to express the variance of the error $\sVar{\sError}$ as:%
\begin{eqnarray}
\sVar{\sError} \approx 
	\left( \sVar{\sLightContribution}+\sEV{\sLightContribution}^2 \right)
	\left( \frac{\sProbabilityR}{\sR}\sConstantKoneD \right)
	\left( \frac{\sProbabilityT}{\sT}\sConstantKt \right), 
	\label{eq:tde_variance_j}
\end{eqnarray}%
where $\sConstantKoneD$ and $\sConstantKt$ are kernel-dependent constants. The last term can be neglected by assuming that the kernels cover small areas in their respective domains, which effectively means that $\sConstantKoneD \gg \sProbabilityR$ and $\sConstantKt \gg \sProbabilityT$. Equation~\eqref{eq:tde_variance_j} shows that for transient density estimation, the variance $\sVar{\sError}$ is inversely proportional to $\sR\sT$.%

\paragraph{Bias} Bias at each iteration $j$ is defined as the expected value of the error $\sEV{\sError_j}$ as
\begin{eqnarray}
\sEV{\sError_j} & = &
		\sEV{
		\sum_{i=1}^M
			\sKoneD \ 
			\sKt \ 
			\sLightContribution -
		\Lo} \nonumber \\
	 & = & \sEV{\sKoneD} \ 
	 	\sEV{\sKt} \ 
	 	\sEV{\sLightContribution } - \Lo. \nonumber
\label{eq:ev_err}
\end{eqnarray}%
Using a second-order expansion of $\sProbabilityT$ and $\sProbabilityR$, instead of the zero$^{th}$-order used when modeling variance, we can express the expected value of $\sKt$ as~\cite{Jarabo2014}%
\begin{eqnarray}
\sEV{\sKt} \approx
	\sProbabilityT + \sT^2
		\int_\Real \sKtCanonical(\sKVarChange)
		\Order{\|\sKVarChange\|^2} \diff\sKVarChange =
	\sProbabilityT + \sT^2 \sConstantSecondKt, \nonumber \\
	& \label{eq:tde_ev_kt_exp}
\end{eqnarray}
while the expected value of $\sKoneD$ is~\cite{jarosz2011progressive}%
\begin{eqnarray}
\sEV{\sKoneD} \approx
	\sProbabilityR +
	\sR \int_{\Real^2}\sKoneDCanonical(\sKVarChange)
	\Order{\|\sKVarChange\|^2} \diff\sKVarChange = 
	\sProbabilityR + \sR \sConstantSecondKoneD, \nonumber \\
	& \label{eq:tppm_ev_kr_exp}
\end{eqnarray}
where $\sConstantSecondKt$ and $\sConstantSecondKoneD$ are constants dependent on the higher-order derivatives of the spatio-temporal light distribution. 
Using \eqref{eq:tde_ev_kt_exp} and \eqref{eq:tppm_ev_kr_exp}, and $\Lo=\sProbabilityR\sProbabilityT\sEV{\sLightContribution}$ we finally compute \sEV{\sError_j} for iteration $j$ as
\begin{eqnarray}
\sEV{\sError_j} & \approx &
	(\sProbabilityR + \sR^2 \sConstantSecondKoneD)(\sProbabilityT +
	\sT^2 \sConstantSecondKt)\sEV{\sLightContribution} -  \sProbabilityR\sProbabilityT\sEV{\sLightContribution }  \nonumber \\
	& = & \sEV{\sLightContribution }(\sProbabilityR\sT^2 \sConstantSecondKt +
	\sProbabilityT\sR^2\sConstantSecondKoneD + \sT^2 \sConstantSecondKt\sR^2\sConstantSecondKoneD).
	\label{eq:tppm_ev_err_pixel_estimate}
\end{eqnarray}%

\section{Convergence Analysis of Progressive Transient Photon Beams}
\label{sec:app_convergence}
Based on the expressions for $\sVar{\sError}$ and $\sEV{\sError_j}$ defined above (Equations~\eqref{eq:tde_variance_j} and \eqref{eq:tppm_ev_err_pixel_estimate}), we can know derive the asymptotic behaviour of Equation~\eqref{eq:amse}. For that, we will compute the variance $\sVar{\estimate{\sPixel}_n}$ and bias $\sEV{\sError_n}$ after $n$ iterations. 

\paragraph{Variance} Assuming that the random variables $\sEyeContribution$ and  $\sError_j$ are independent, we model the variance of the estimator $\sVar{\estimate{\sPixel}_n}$ in Equation~\eqref{eq:app_variance_n} as~\cite{knaus2011progressive}:%
\begin{eqnarray}
\sVar{\estimate{\sPixel}_n} & = &
	\frac{1}{n} \sVar{\sEyeContribution\Lo} + \frac{1}{n^2} \sum_{j=1}^n\sVar{\sEyeContribution\sError_j}  \\ 
	& = & \frac{1}{n} \sVar{\sEyeContribution\Lo} + \sVar{\sEyeContribution}
	\frac{1}{n^2} \sum_{j=1}^n \sVar{\sError_j} +\nonumber \\
	& & \sEV{\sEyeContribution}^2 \frac{1}{n^2} \sum_{j=1}^n\sVar{\sError_j} +
	\sVar{\sEyeContribution} \frac{1}{n^2} \sum_{j=1}^n\sEV{\sError_j}^2 . \nonumber
	\label{eq:tppm_var_pixel_estimate}
\end{eqnarray}
Following~\cite{kaplanyan2013adaptive}, we can approximate $\sVar{\sError_n}$ as a function of the variance at the first iteration $\sVar{\sError_1}$ as:%
\begin{equation}
\sVar{\sError_n} \approx \frac{\sVar{\sError_1}}{(2-\alpha)n^\alpha} = \Order{n^{-\alpha}}. 
\end{equation}
Finally, by applying $\sVar{\sError_n}$ and asypmtotic simplifications,  we can formulate $\sVar{\estimate{\sPixel}_n}$~\eqref{eq:tppm_var_pixel_estimate} as:%
\begin{eqnarray}
\sVar{\estimate{\sPixel}_n} &\approx&
	\frac{1}{n} \sVar{\sEyeContribution\Lo} +
	\sEV{\sEyeContribution}^2\sVar{\sError_n} \nonumber \\ 
	&\approx& \frac{1}{n} \sVar{\sEyeContribution\Lo} +
	\frac{\sVar{\sError_1}}{(2-\alpha)n^\alpha}  \nonumber \\
	&=& \Order{n^{-1}} + \Order{n^{-\alpha}} = \Order{n^{-\alpha}}.
	\label{eq:tppm_variance_n_asymptotic}
\end{eqnarray}

\paragraph{Bias} \label{sec:var_exp_err_pix_estimate} The expected value of  the error $\sEV{\sError_n}$ is modeled in Equation~\eqref{eq:tppm_ev_pixel_estimate} as a function of the averaged bias introduced at each iteration $\sEV{\sError_j}$~\eqref{eq:tppm_ev_err_pixel_estimate}. Computing the kernels' bandwidth $\sT_{j}$ and $\sR_{j}$ at iteration $j$ by expanding Equation~\eqref{eq:tppm_alpha} as a function of their initial value by we get%
\begin{eqnarray}
\sT_{j} &=&
	\sT_{1}(j \ \alpha \ \Beta(\alpha,j))^{-\scaleParT},
	\label{eq:tppm_Tj} \\
\sR_{j} &=&
	\sR_{1}(j \ \alpha \ \Beta(\alpha,j))^{-\scaleParR},
	\label{eq:tppm_Rj}
\end{eqnarray}%
where $\Beta(x,y)$ is the Beta function. Using \eqref{eq:tppm_Tj} and \eqref{eq:tppm_Rj} in Equation~\eqref{eq:tppm_ev_err_pixel_estimate} we can express \sEV{\sError_j} as a function of the initial kernel bandwidths
\begin{eqnarray}
\sEV{\sError_j} &=&
	\sEV{\sLightContribution }\sProbabilityR
	\sConstantSecondKt \sT_{1}^2\order{j^{1-\alpha}}^{-2\scaleParT}  \nonumber \\ 
	& & + \sEV{\sLightContribution}\sProbabilityT
	\sConstantSecondKoneD\sR_{1}^2\order{j^{1-\alpha}}^{-2\scaleParR} \nonumber \\
	& & + \sEV{\sLightContribution }\sConstantSecondKt
	\sConstantSecondKoneD \sT_{1}^2\sR_{1}^2\order{j^{1-\alpha}}^{-2(\scaleParT+\scaleParR)}.
	\label{eq:tppm_ev_err_kernel_j1_asymptotic}
\end{eqnarray}

Finally, we use $\sum_{j=1}^n \order{j^{x}} = n \ \Order{n^x}$ to plug Equation~\eqref{eq:tppm_ev_err_kernel_j1_asymptotic} into Equation~\eqref{eq:tde_bias_n} to get the asymptotic behavior of $\sEV{\sError_n}$ in transient progressive photon beams:
\begin{eqnarray}
	\sEV{\sError_n} =
		\Order{n^{1-\alpha}}^{-2\scaleParT} +
		\Order{n^{1-\alpha}}^{-2\scaleParR} +
		\Order{n^{1-\alpha}}^{-2 (\scaleParT +
		\scaleParR)}, \nonumber
\end{eqnarray}
\noindent which, by using the equality $\scaleParR=1-\scaleParT$, becomes:
\begin{eqnarray}%
\sEV{\sError_n} &=&
	\Order{n^{1-\alpha}}^{-2\scaleParT}+\Order{n^{1-\alpha}}^{2 \scaleParT-2}+\Order{n^{1-\alpha}}^{-2} \nonumber \\
	&=& \Order{n^{1-\alpha}}^{-2\scaleParT}+\Order{n^{1-\alpha}}^{2 \scaleParT-2}.
	\label{eq:tppm_bias_n_asymptotic}
\end{eqnarray}

\section{Minimizing Asymptotic Mean Squared Error}%
\label{sec:minimizing_amse}
Using the asymptotic expression for variance and bias in Equations \eqref{eq:tppm_variance_n_asymptotic} and \eqref{eq:tppm_bias_n_asymptotic}, we can express the AMSE~\eqref{eq:amse} as%
\begin{eqnarray}
AMSE(\estimate{\sPixel}_n) =
	\Order{n^{-\alpha}} + 
		\left(\Order{n^{1-\alpha}}^{-2\scaleParT} +
		\Order{n^{1-\alpha}}^{2 \scaleParT-2}\right)^2. \nonumber \\
		 & \label{eq:AMSE_asymptotic}
\label{eq:tppm_amse}
\end{eqnarray}%
\noindent which is a function of the parameters $\alpha$ and $\scaleParT$. Given that the variance is independent of $\scaleParT$, we first obtain the optimal value for this parameter that yields the highest convergence rate of the bias \sEV{\sError_n}. We differenciate Equation~\eqref{eq:tppm_bias_n_asymptotic}, apply asymptotic simplifications and equating to zero, we obtain the optimal value $\scaleParT=1/2$. By plugging this value in Equation~\eqref{eq:AMSE_asymptotic}, we obtain:
\begin{equation}
AMSE(\estimate{\sPixel}_n) = \Order{n^{-\alpha}} + \Order{n^{-2 (1-\alpha)}}.
	\label{eq:tppm_amse_beta}
\end{equation}%
Finally, by finding the minimum again with respect to $\alpha$ we get the optimal parameter $\alpha=2/3$, which results in the optimal convergence rate of the AMSE for our transient progressive photon beams as
\begin{equation}
AMSE(\estimate{\sPixel}_n) = \Order{n^{-\frac{2}{3}}}+\Order{n^{-2(1-\frac{2}{3})}} = \Order{n^{-\frac{2}{3}}}.
	\label{eq:tppm_optimal_amse}
\end{equation}
}


\bibliographystyle{eg-alpha-doi}

{\bibliography{bibliography}}

\newpage

\end{document}